%% file: main.tex
\documentclass[twocolumn]{aastex631}

\usepackage{bm,amsmath}
\usepackage{comment}
\usepackage{url}
\usepackage{soul}

\begin{document}

\title{The Cosmic-ray Knee as a Local Signature of Nearby PeVatrons}

\author[0000-0002-5387-8138]{Ke Fang}
\affiliation{Department of Physics, Wisconsin IceCube Particle Astrophysics Center, University of Wisconsin, Madison, WI, 53706}

\author[0000-0001-6224-2417]{Francis Halzen}
\affiliation{Department of Physics, Wisconsin IceCube Particle Astrophysics Center, University of Wisconsin, Madison, WI, 53706}

\date{\today}

\begin{abstract}

A ``knee" in the cosmic-ray spectrum, characterized by a sudden steepening of the spectral shape at $\sim 4$~PeV, may be interpreted either as a global feature of Galactic cosmic rays or as a local signature. In the former scenario,  cosmic-ray spectra throughout  the Galaxy would be similar to that observed in the solar neighborhood, and the knee would be a common feature of the cosmic-ray sea. In the latter scenario, the PeV cosmic-ray flux varies across the Galactic disk, and the knee is dominantly contributed by a small number of nearby sources. By simulating cosmic-ray propagation in the Galactic magnetic field and interstellar medium, we show that the two scenarios correspond to different regimes of the birth rate of PeV proton accelerators and depend on the presence  of powerful  nearby sources. By comparison with both cosmic-ray and gamma-ray observations, 
we find that a local knee would be best explained by sources located at distances of order $\sim1$~kpc and with ages in the range 0.1-1~Myr, with the Cygnus Cocoon being a particularly promising candidate. 
\end{abstract}


\section{introduction}

The knee of the cosmic-ray spectrum at around 4~PeV \citep{Hoerandel:2002yg} and the related sharp softening of the proton spectrum at 3~PeV  \citep{LHAASOKnee} indicates that certain Galactic sources, commonly referred to as PeVatrons, accelerate protons up to multi-PeV energies. The spectrum  of the accelerated particles, $dN/dE\propto E^{-s}$, needs to be relatively hard, with $s \approx 2.1$-$2.4$ \citep{2012JCAP...01..010B}, so that after  propagation the observed spectral index reaches $s_{\rm obs}\sim 2.7$ up to the knee.

A direct search for PeVatrons in the ultrahigh-energy (UHE; $>100$~TeV) gamma-ray sky has recently become possible with gamma-ray air shower observatories. The 1LHAASO catalog reports 43 UHE gamma-ray sources \citep{1LHAASO}. However, most of these sources exhibit softer spectra with $s\gtrsim 3$ at 50~TeV, and about half of them are associated with pulsars,   whose TeV gamma-ray emission is accommodated by inverse Compton scattering by electrons. Only a handful of promising PeVatron candidates have been identified, including the Galactic center \citep{HESS:2016pst}, star-forming region Cygnus Cocoon \citep{Abeysekara:2021yum,LHAASO:2023uhj}, the supernova remnant G106.3+2.7 \citep{TibetASg:2021kgt, Fang:2022uge} and the microquasar V4641~Sgr \citep{Alfaro:2024cjd, LHAASO:2024psv}.


The scarcity of PeVatrons in the UHE gamma-ray sky suggests that PeVatrons may be rare compared to the supernova rate, in which case PeV cosmic rays originate from old sources that are no longer active \citep{2020APh...12302492C}. Alternatively, they may be inefficient gamma-ray emitters so that the majority lie below the detection threshold; see example \citet{Fang:2024wmf}.

While GeV--TeV cosmic rays form a smoothly distributed ``cosmic-ray sea'' in the Galactic disk, the distribution may be less homogeneous at energies near the knee. This can arise from two effects. First, as implied by UHE gamma-ray observations, the population of PeVatrons could be small. Second, PeV cosmic rays have much shorter residence times in the Galaxy than GeV-TeV cosmic rays, making their spatial distribution more dependent on the recent history of sources \citep{Kaci:2024wra, Stall:2025ggd}. As a consequence, localized ``cosmic-ray islands'' (sometimes referred to in the literature as local cosmic-ray bubbles) may form, characterized by enhanced cosmic-ray density and upward fluctuations in the cosmic-ray spectrum at PeV energies.

The possibility that few nearby bright sources contribute to the cosmic-ray knee has been investigated in the literature based on both cosmic-ray spectral and anisotropy measurements. The rapid rise and steepening of the cosmic-ray spectrum at the knee was interpreted as the additional contribution from a single source \citep{Erlykin_1997} or from several nearby sources \citep{2013APh....50...33S, Evoli:2021ugn} superimposed on a smoothly steepening background component. Consistent with this picture, the large-scale cosmic-ray anisotropy exhibits an amplitude and phase that are essentially independent of energy between 1 and 100~TeV, while at higher energies the amplitude increases monotonically up to $\sim 30$~PeV where the phase flips compared to lower energies \citep{IceCube:2024pnx}. This trend is interpreted as the transition from an anisotropy controlled by the local magnetic field below 100~TeV to the dominance of nearby individual sources at PeV energies \citep{Erlykin:2006ri, Ahlers:2016njd}.

In this paper, we will show that the two phenomena, namely the presence of cosmic-ray islands and a significant contribution from nearby sources, are correlated and naturally arise when PeVatrons have a low birth rate (Section~\ref{sec:birthrates}). In light of the latest air shower gamma-ray observations, we examine the properties of a potential knee-producing source in Section~\ref{sec:testSource}. Based on these results, we speculate in Section~\ref{sec:cygnus} that the Cygnus Cocoon, a promising PeVatron in the Northern sky, may be among the nearby sources that make an important contribution to the knee.

\section{PeV particles in the Galaxy: islands in the cosmic-ray sea}\label{sec:birthrates}

\begin{figure*}
    \centering
   \includegraphics[width=1.06\textwidth]{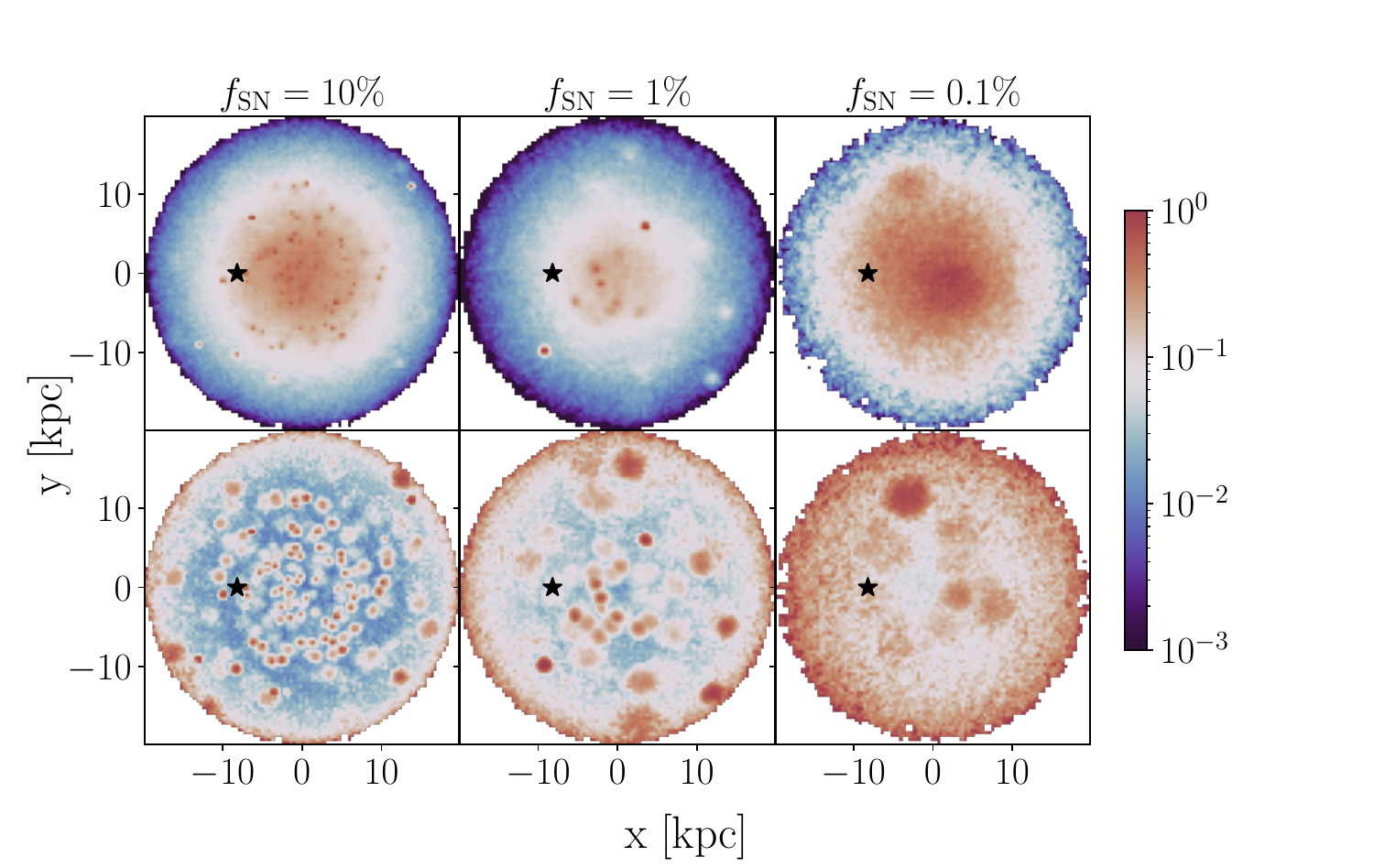}
   \caption{
    \label{fig:1PeV} Counts of 1~PeV cosmic rays normalized to the maximum pixel (top row) and the maximum fraction originating from a single source (bottom row) for locations across the disk with $|z|<0.17$~kpc, sampled in spacial bins of 0.4~kpc in the x and y directions.  Sources are assumed to follow a pulsar distribution in space and uniform distribution in time, and steadily emit cosmic rays during a lifetime of 100~kyr. The left, middle, and right panels correspond to source birth rates of 10\%, 1\%, and 0.1\% of a baseline, supernova-like rate of $0.01\,\rm yr^{-1}$, respectively. The black star marks the position of the Sun.    } 
\end{figure*}

We first investigate how the homogeneity of PeV cosmic-ray density in the Galactic disk depends on distribution and evolution history of the PeVatrons. Because their population remain largely unknown, we treat their birthrate   as a free parameter, normalized to the supernova rate of $0.01\,\rm yr^{-1}$ \citep{1994ApJS...92..487T}.
The rate is parametrized as a fraction $f_{\rm SN}$ of the supernova rate, with representative values $f_{\rm SN}=$  $10\%$, $1\%$, and $0.1\%$, corresponding to $N_s=10^5$, $10^4$ and $10^3$ sources over the past 100~Myr. The generic population of PeVatrons are assumed to inject 1 PeV protons with identical  rates  and durations of $\Delta t = $100~kyr. These simplifying assumptions reduce the number of parameters which are however sufficient to demonstrate the important features of our proposal.

In the simulation, we randomly generate $N_s$ sources with positions in the Galactic disk following the pulsar distribution, which is known to correlate with star formation regions \citep{2006ApJ...643..332F, 2012JCAP...01..010B}, and with ages drawn uniformly between 0 and 100 Myr. We propagate the injected protons in the Galactic magnetic field using an anisotropic diffusion model, with a parallel diffusion coefficient $D_\parallel = 6.1\times 10^{28} \left( {R}/{4\,\rm GV}\right)^{1/3}\,\rm cm^2\,s^{-1}$ \citep{Strong:1998pw}, where $R = E/Z$ is the rigidity, and a perpendicular diffusion coefficient $D_\perp = 0.1 D_\parallel$. Details of the numerical setup are described in Appendix~\ref{appendix:setup}.

Figure~\ref{fig:1PeV} presents the cosmic-ray density in the Galactic disk in the top row. In all cases, the PeV cosmic-ray density in the disk is smooth overall. The inner Galaxy exhibits a higher density than the outer Galaxy as a result of its higher star-formation rate. When the source rate is $\gtrsim 1\%$ supernova rate, most regions of the disk are dominated by the cosmic-ray sea, that is, the diffuse background of cosmic rays consisting of particles injected elsewhere and propagate to the location in the disk. However, in the immediate vicinity of young, active sources, PeV cosmic rays have not yet diffused far from their birth sites, producing localized overdensities that appear as bright spots. 
For  the $f_{\rm SN}= 10\%$ and $1\%$ scenarios, the model predicts tens to hundreds of such sources with ages younger than a few hundred kyr.
In contrast when the PeVatron fraction is low, e.g $f_{\rm SN}= 0.1\%$, the cosmic-ray density is primarily determined by a handful of recent sources with ages of a few Myr, resulting in a more spatially variable distribution. 

The bottom row of Figure~\ref{fig:1PeV} shows the contribution from the leading source, defined as the source that provides the largest fraction of the total PeV cosmic-ray flux at a given position in the disk today. The fractional contribution from a single source exhibits a dependence on the source rate that has been discussed for the top-row distributions. When sources are abundant, the cosmic-ray flux at a given location is dominated by the background, with individual sources typically  contributing at no more than the percent level, except in the vicinity of  a young or active source. The spatial extent of regions dominated by a single source is small. This is because a large source population produces a smooth and substantial background, and individual sources rarely stand out. In contrast, when sources are rare, the region influenced by an individual source becomes much larger. Consequently, a greater fraction of the Galactic disk can have a significant contribution from a single source. For positions in the disk within a height $|z| < 0.17$~kpc, we find average leading-source contribution of $7.5\%$, $11.3\%$, and $20.2\%$ for $f_{\rm SN} = 10\%$, 1\%, and 0.1\% scenarios, respectively. 

Overall the results suggest that the knee may be a local effect, indicating that we reside in a cosmic-ray island rather than a cosmic-ray sea, if recent active sources are located nearby. The relative contribution from an individual source is enhanced when the PeVatron population is sparse.

We have assumed a constant source duration of 100~kyr,  which lies between the characteristic lifetimes of several relevant source classes such as supernova remnants ($\sim 10-100$~kyr; \citealp{2005JPhG...31R..95H}),   microquasar jets (tens to hundreds of kyr; \citealp{2005Natur.436..819G,2011MNRAS.414.2838G}), and star-forming episodes ($\leq1$~Myr to tens of Myr; \citealp{2005ApJ...630..250K}). We show the effects of source duration in Appendix~\ref{appendix:sourceLifetime}. In general, a longer source duration would lead to a smoother cosmic-ray density and lower single-source contribution, as the source activity would be more uniformly distributed in time. Conversely, a shorter source duration would result in a less smooth density and a larger contribution from individual sources, especially when the source birth rate is low.

\section{The characteristics of a knee-producing source} \label{sec:testSource}

We next investigate the properties of a nearby source that could play a significant role in shaping the cosmic-ray knee. As a fiducial configuration, we place the source at a distance of 1~kpc in the direction of the Galactic center. 
We consider four source ages: $t_{\rm age}=$~10~kyr, 100~kyr, 1~Myr, and 10~Myr. To enable a fair comparison among these cases, we inject the same proton spectrum, which is a power-law spectrum with a cutoff at the knee's energy, $dN/dE\propto E^{-2}\exp(-E/3\,{\rm PeV})$, and fix the total proton energy, integrated from the proton rest mass to 100~PeV, to $W_p = 10^{51}$~erg. The proton emission is assumed to be continuous at a constant rate over the entire source duration. For active sources, this source duration is taken to be equal to the source age, $\Delta t = t_{\rm age}$. Relic sources, which are no longer producing cosmic rays with $\Delta t < t_{\rm age}$,  are discussed in Appendix~\ref{appendix:sourceLifetime}.

\begin{figure}
    \centering
   \includegraphics[width=0.49\textwidth]{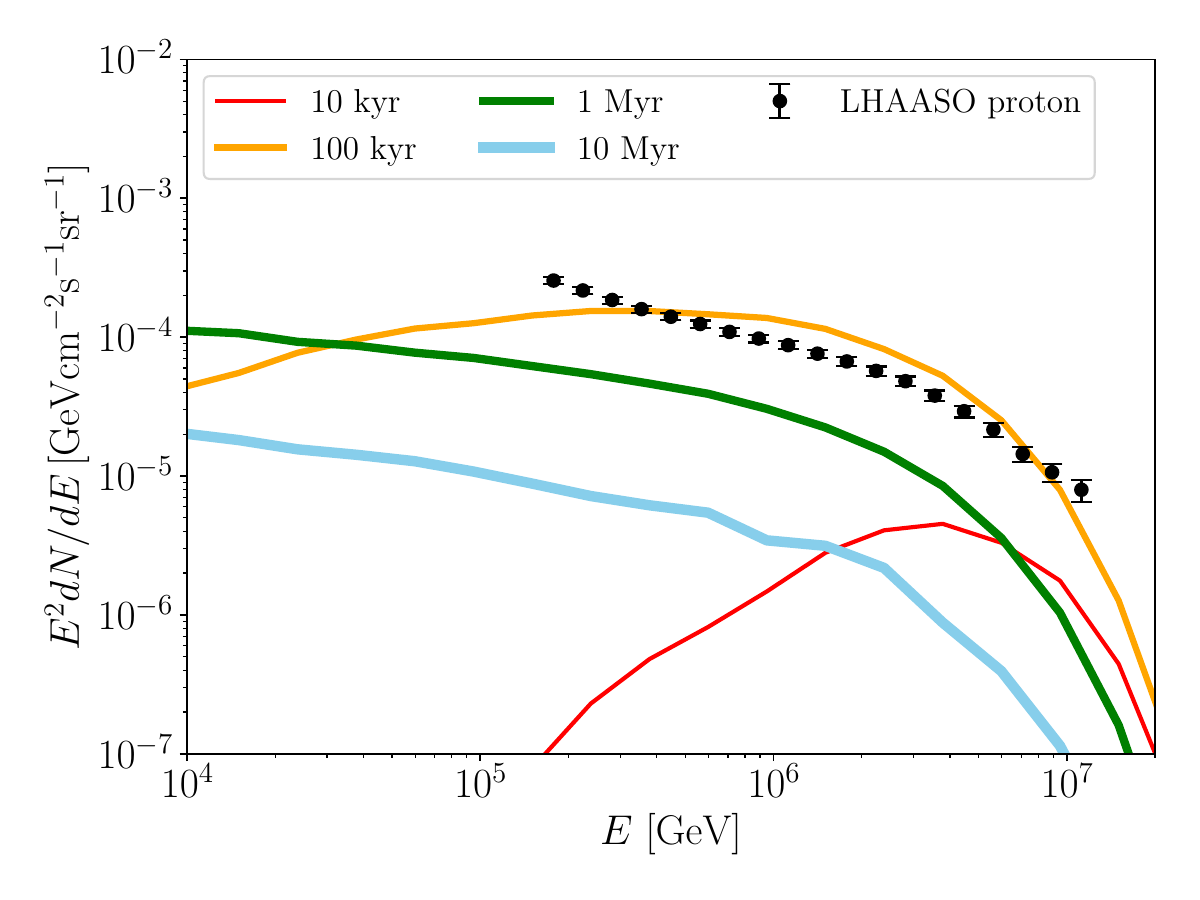}
    \caption{
    \label{fig:spectrum_source} Proton spectrum at the sun's position from a fiducial source at 1~kpc away in the direction of the Galactic center, with an age of 10~kyr (red), 100~kyr (orange), 1~Myr (green), and 10~Myr (blue). The cosmic rays are injected following an $dN/dE\propto E^{-2}\exp(-E/3\,{\rm PeV})$ spectrum with a total energy of $10^{51}$~erg from the proton rest mass to 100~PeV. The black data points show the measurement of the proton spectrum by LHAASO \citep{LHAASOKnee}. } 
\end{figure}

\begin{figure*} 
    \centering
   \includegraphics[width=0.99\textwidth]{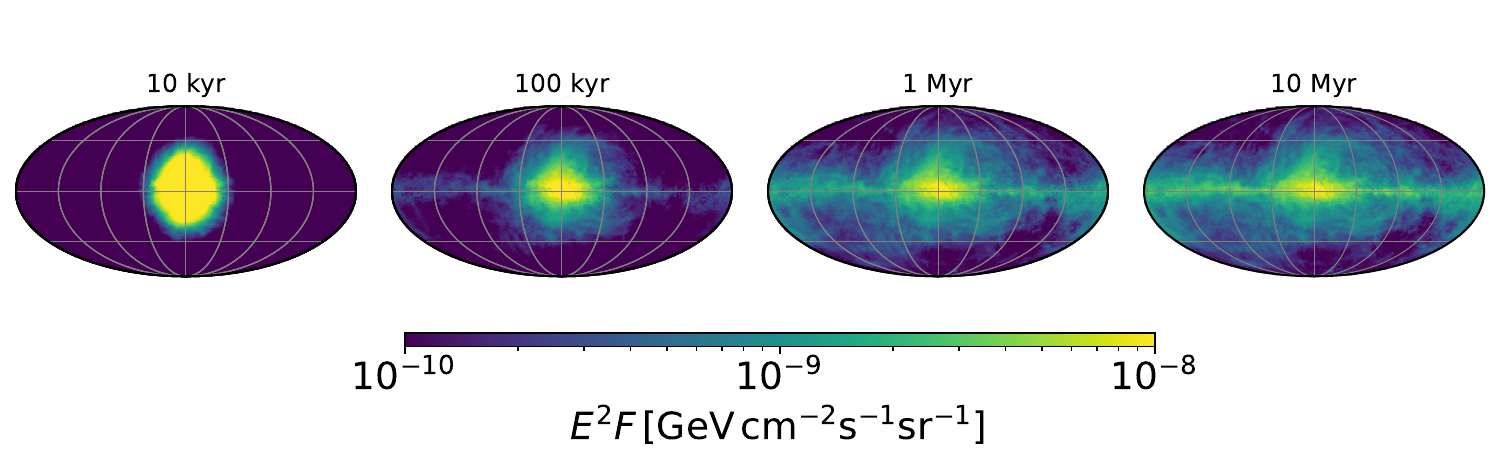}
    \caption{
    \label{fig:skymap_source} Gamma-ray emission at 100~TeV produced by cosmic rays diffusing from test sources aged 10~kyr to 10~Myr, through interactions with neutral hydrogen. The proton spectrum is assumed to follow $dN/dE_p\propto E_p^{-2}\exp(-E/3\,{\rm PeV})$ as in Figure~\ref{fig:spectrum_source} and is normalized to the LHAASO data point at 3~PeV. The total cosmic-ray injection energy above the proton rest mass in the four cases are $10.5$, $0.7$, $3.9$ and $29.5$ $\times 10^{51}$~erg, respectively.  } 
\end{figure*}

\subsection{Cosmic-ray emission}\label{subsec:cosmicRay}

Figure~\ref{fig:spectrum_source} shows the spectrum of the cosmic rays from the fiducial source observed at the solar position. We also present the spatial distribution of the spectral index of cosmic rays from these sources in Figure~\ref{fig:index_source} in  Appendix~\ref{appendix:index}.

For young sources ($t_{\rm age} \sim 10$~kyr), the highest-energy cosmic rays just reach the observer while the lower-energy particles remain closer to the source, yielding a hard local spectrum. For middle-aged sources ($t_{\rm age} \sim 100$~kyr), cosmic rays propagate further and the lower-energy cosmic rays catch up, bringing the local spectrum closer to the injection spectrum. A source in this age group efficiently contributes to the PeV cosmic-ray flux because its PeV particles are arriving at the observer after traveling for $t \sim d^2/(4D_\parallel) = 20\,\mathrm{kyr}\,(d/1\,\mathrm{kpc})^2\,(E/1\,\mathrm{PeV})^{-1/3}$. 
On even longer timescales ($t_{\rm age} \gtrsim 1$~Myr), the faster escape of higher-energy cosmic rays leads to an observed spectrum that is softer than the injection spectrum. For old sources, most cosmic rays injected in its early times have left the Galaxy, substantially reducing the observed flux.


Figure~\ref{fig:spectrum_source} suggests that a nearby source that may dominantly contribute to the knee must have a significant proton energy, close to or exceeding the canonical supernova budget. This energy requirement disfavors sources that are too young ($\ll 10$~kyr) or too old ($\gg 1$~Myr) as a major contributor, unless the source lies much closer than 1~kpc or, is more powerful than a supernova such as accretion onto the supermassive black hole at the Galactic center. Instead, it favors middle-aged or recurrent engines, such as star-forming regions or microquasar jets.

\subsection{Gamma-ray emission}\label{subsec:gamma}

We next compute the diffuse gamma-ray emission produced by cosmic rays originating from the fiducial sources. The calculation of the secondary emission is described in Appendix~\ref{appendix:secondary}.

Figure~\ref{fig:skymap_source} shows the map of gamma rays produced by cosmic-ray interactions with the interstellar medium (ISM) for the four source ages. In the 10~kyr case, particles have not propagated far, producing a bright, very extended, roughly spherical halo centered on the source. Such an object would outshine the Galactic plane and be easily identifiable by wide-field gamma-ray observatories. Conversely, the absence of an extremely extended, spherical source in the UHE gamma-ray sky disfavors a scenario where a young source powers the knee. At 100~kyr, the halo persists but becomes more elongated and blends into more diffuse-like emission along the Galactic plane as particles travel farther. For ages $\gtrsim1$~Myr, particles have traversed large distances in the disk while leaking out vertically, yielding emission that is predominantly diffuse-like and difficult to distinguish from that produced by a cosmic-ray sea. We here assumed that the source remains active over its entire age. For a relic source that has been inactive for a long time, the associated gamma-ray emission would appear fully diffuse and would no longer reveal the presence of the source. 

Both the cosmic-ray flux and the gamma-ray map indicate that a source at $d\sim1$~kpc with an age $t\sim0.1$–$1$~Myr represents the most plausible PeVatron contributing substantially to the knee, without requiring an extreme energy budget or producing an obviously overbright and highly extended gamma-ray halo. This conclusion differs from traditional assumptions in the literature, which are based on cosmic-ray observations alone and often favor young supernova remnants at sub-kiloparsec distances. The signature of such a local PeVatron would be UHE gamma-ray emission over a highly extended region surrounding a more compact core where cosmic rays interact with the source medium.

\section{Potential sources}\label{sec:cygnus}

The obvious question is whether there are sources in the UHE gamma-ray sky with the characteristics identified in Section~\ref{sec:testSource} that can be associated with the appearance of the knee in the Galactic cosmic-ray spectrum. 

Among the UHE gamma-ray sources in the 1LHAASO catalog \citep{1LHAASO}, which covers most of the Northern sky, three lie within 2~kpc and are not associated with a pulsar: 1LHAASO J2228+6100u, (SNR G106.3+2.7), 1LHAASO J2031+4052u* (LHAASO J2032+4102 in the Cygnus-X region), and 1LHAASO J1850-0004u* (SNR G031.5-0.6) \footnote{We note that sources associated with pulsars could still accelerate hadronic cosmic rays, even though their gamma-ray emission is often well described by inverse-Compton radiation. In addition, a substantial fraction of the UHE sources lack firm associations and therefore reliable distance estimates.}. Of these, only LHAASO J2032+4102 is middle-aged. The Cygnus-X region also exhibits UHE gamma-ray emission with energies exceeding 1~PeV extending over at least 100~deg$^2$ \citep{LHAASO:2023uhj}. 

Cygnus-X is an active star-forming complex. It hosts the largest star-forming region in the solar neighborhood and contains numerous stellar associations with ages of a few Myr and possible past supernova events; see e.g., \citealp{2020A&A...642A.168B}. An extended gamma-ray source with emission tracing the infrared emission by gas and dust in the region has been revealed by {\it Fermi}-LAT \citep{Ackermann:2011lfa, Astiasarain:2023zal} and HAWC  
\citep{Abeysekara:2021yum}, known as the Cygnus Cocoon. Its spectrum, morphology, and lack of an X-ray counterpart \citep{Guevel:2022esw} point to a hadronic origin, suggesting proton acceleration by the stellar winds of young stars \citep{Ackermann:2011lfa} or sources embedded in the complex, such as microquasar  Cygnus~X-3 \citep{LhaasoCollaboration:2025jwd} or a supernova remnant \citep{2025arXiv250821644H}. After removing individual gamma-ray sources, LHAASO finds an even larger structure, spreading to a radius of $\sim 6^\circ$, termed the Cygnus Bubble \citep{LHAASO:2023uhj}. 

\begin{figure*}
    \centering
   \includegraphics[width=0.49\textwidth]{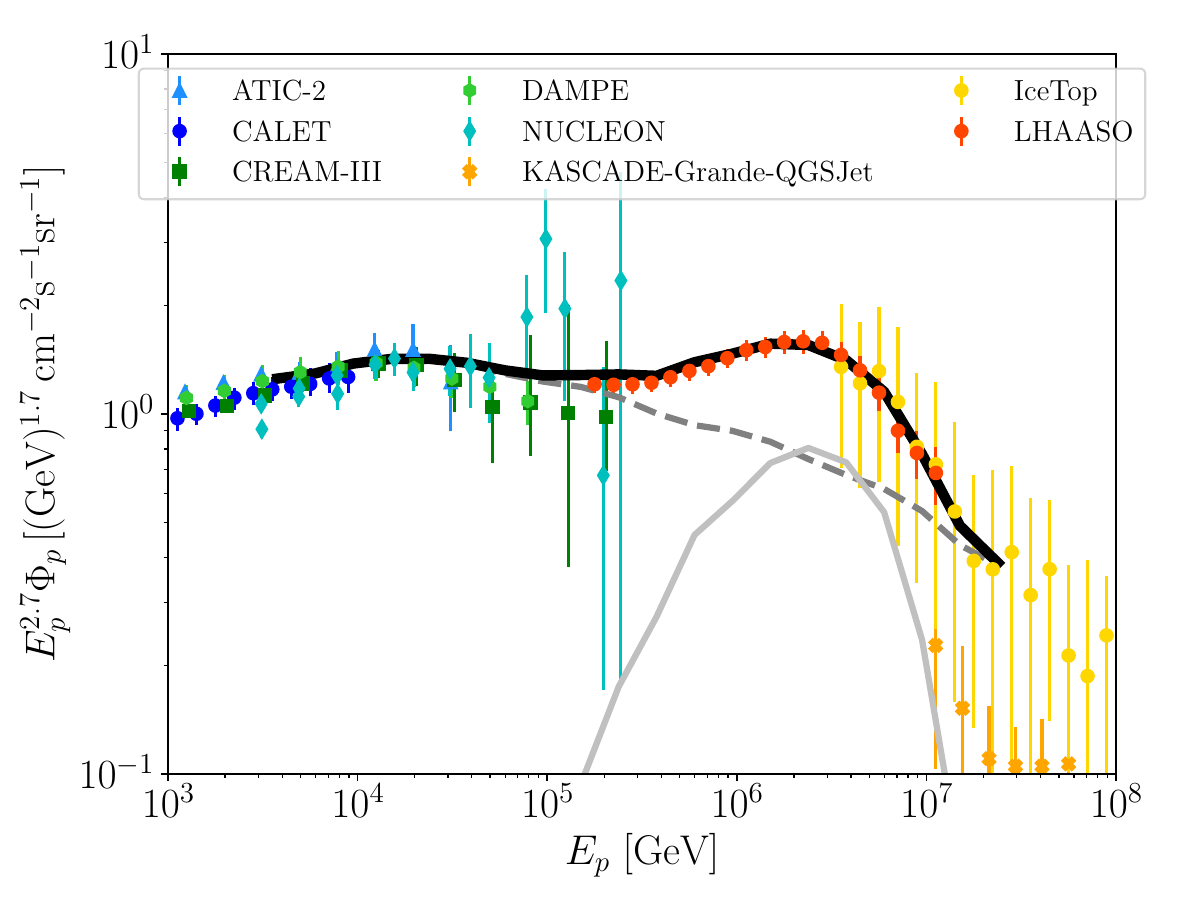}
   \includegraphics[width=0.49\textwidth]{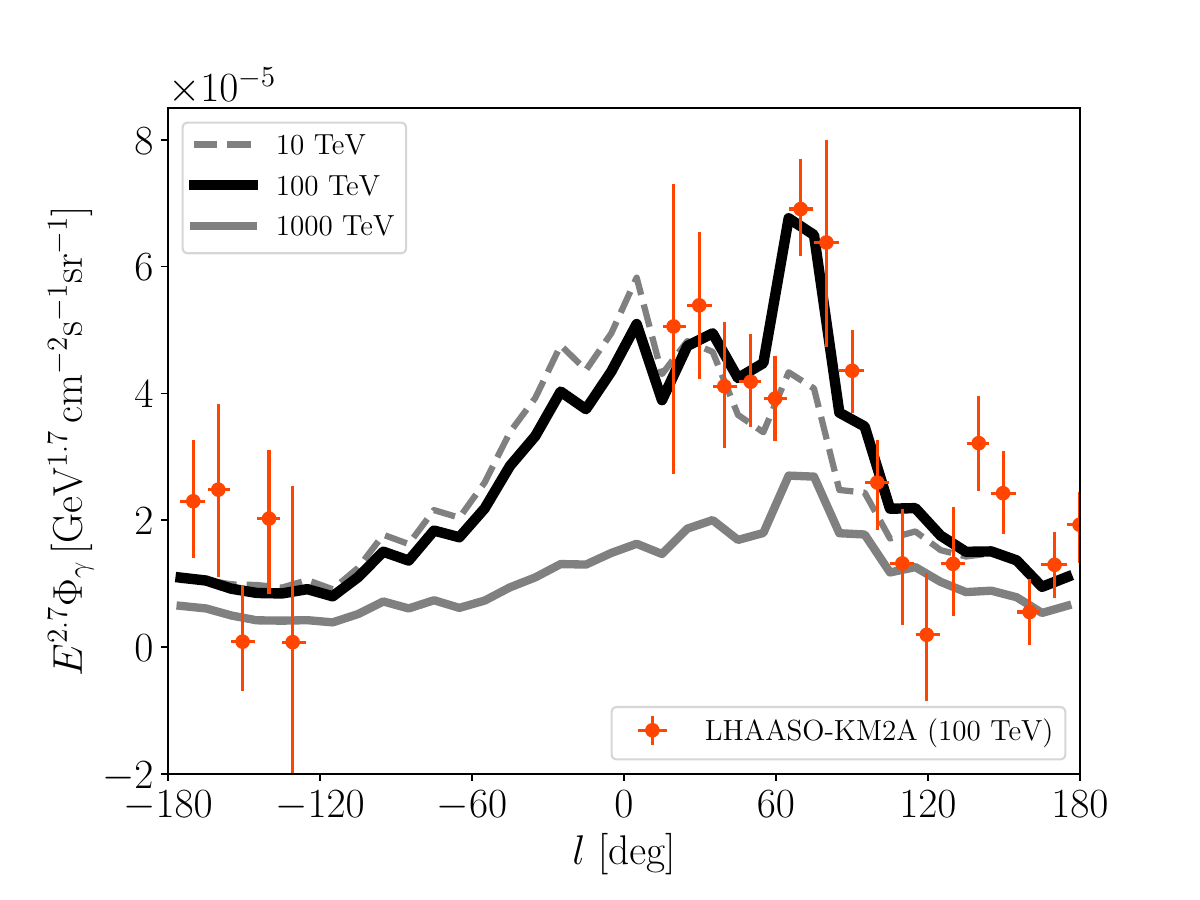}
    \caption{
    \label{fig:cygnus} Cosmic-ray spectrum (left) and longitudinal distribution of the diffuse gamma-ray emission from their interaction with neutral (HI) and molecular hydrogen (H$_2$) (right) from a model where a large population of cosmic-ray sources following the pulsar distribution dominantly contribute to cosmic rays below 100~TeV and a single source, the Cygnus Cocoon, dominantly contributes to the knee.      } 
\end{figure*}

The substantial UHE extent of the Cygnus Bubble with the Cygnus Cocoon as its bright core, and possible supernova or microquasar jet activity within the past $\sim$0.1–1~Myr are all consistent with the properties derived above for a plausible local contributor to the knee. Below, we explore a possible association of the Cygnus region with a cosmic ray island producing the knee.

To keep the model as generic as possible, we assume that sources of cosmic rays below $\sim 100$~TeV yield a broadly diffuse cosmic-ray spectrum across the disk. Rather than modeling this population in detail, we adopt an injection spectrum with a super-exponential cutoff, $dN/dE\propto E^{-2.37} \exp[-(E/1 \,{\rm PeV})^{0.2}]$. Such a super-exponential cutoff approximately produces an $E^{-3}$ spectrum above 100~TeV that extends into the transition regime. It naturally results from a population of sources with varying maximum energies. For the Cygnus cocoon, at a distance \footnote{Different components of the Cygnux X complex are located at slightly different distances. A related discussion is presented in \citet{2025arXiv250821644H}.}  of 1.4~kpc and Galactic coordinates ($l =79^\circ.62$, $b = 0^\circ.96$), we assume an injection spectrum $dN/dE\propto E^{-2}\exp(-E/E_{\rm max})$. With a source age of 100~kyr, the knee can be reproduced with $E_{\rm max} = 3$~PeV and a total proton energy above the proton rest mass  of $W_p = 8.4\times 10^{50}$~erg. The left panel of Figure~\ref{fig:cygnus} compares this model to the measurements of the local proton spectrum. The model is not fine-tuned, and  similar results are obtained for  source ages of a few hundred kyr and $E_{\rm max}$ in the range of $\sim 3-7$~PeV.

The right panel of Figure~\ref{fig:cygnus} compares the longitudinal distribution of the diffuse gamma-ray emission with the  LHAASO-KM2A measurements. We convert the integrated flux between 63 and 1000 TeV measured by LHAASO to a differential flux at 100~TeV using their best-fit spectral index of $-2.99$ \citep{lhaasoGDE}. To ensure consistency with LHAASO’s methodology for measuring the diffuse gamma-ray emission, we also apply a $6^\circ$ mask centered on the Cygnus Cocoon. The plot shows that cosmic rays in the vicinity of the Cygnus Cocoon account for the significant residual emission near $l = 80^\circ$ while those traveling further away from the source accommodate the diffuse emission in longitudinal band between $l\sim 30^\circ$ and $120^\circ$. The model predicts an even more prominent bump at the position of the knee-producing source at higher energies (the grey solid curve in the right panel of Figure~\ref{fig:cygnus}), where the source contribution to the knee exceeds the background. This prediction may be tested by future observations.

\section{Discussion} \label{sec:discussion}

Motivated by recent cosmic ray and gamma ray measurements by LHAASO in the PeV-energy regime, we revisited the scenario where the knee in the cosmic ray spectrum is significantly contributed by nearby sources, also referred to as the single-source model in literature. Our conclusions may be summarized as follows:

\begin{itemize}
    \item Assuming the standard diffusion paradigm, we find that the PeV cosmic-ray density in the Galactic disk is generally smooth, except when the PeVatron birth rate is $\lesssim 1\%$ of the supernova rate, in which case the distribution becomes dominated by a few recent sources. Regardless of the birth rate, significant upward fluctuations are expected in the immediate vicinity of a young, active source with an age less than a few hundred kyr. In such regions, the contribution from a single source may reach $10\%$ or higher. 
    \item By combining cosmic-ray and gamma-ray observations we find that that the most promising knee-producing sources are located at $\sim 1 $~kpc distances and have ages of  $\sim 0.1-1$~Myr. Younger and closer sources would produce overly bright and extended halos that are inconsistent with TeV-PeV gamma-ray observations, while older and more distant sources would require an extreme energy budget. 
    \item A model invoking a significant contribution from the Cygnus Cocoon to the knee may naturally explain the excess in the gamma-ray Galactic diffuse emission around $l\sim 80^\circ$ observed by KM2A at $\sim 63-1000$~TeV. 
\end{itemize}

Another constraint on the contribution from nearby sources arises from the cosmic-ray anisotropy measurements. In the vicinity of the knee, the cosmic-ray anisotropy has an amplitude of a few times of $10^{-3}$ \citep{IceCube:2024pnx}. A nearby source would result in a dipole oriented towards the source at a distance $d$ with an amplitude of $\delta\approx  3D(E) /c \, (\nabla n/   n) \sim   D(E) / c d =$ a few $\times 10^{-3} (E/1\,{\rm PeV})^{1/3} (d/1\,\rm kpc)^{-1}$, for proton energy $E$ and assuming a Kolmogorov diffusion $D(E)\approx10^{28}(E / 3\,\rm GeV)^{1/3}$ \citep{2012PhRvL.109g1101G}. The large scale anisotropy is further impacted by the directions of the other nearby sources, the presence of helium and heavier nuclei at the knee energy, as well as the anisotropic diffusion along  the local  magnetic field \citep{2012JCAP...01..011B,2014Sci...343..988S, Evoli:2021ugn}. Notably, the direction of the Cygnus Cocoon ($\alpha\approx 307^\circ.7$; \citealp{Abeysekara:2021yum}) is close to the phase of the large-scale anisotropy measured by Tibet, IceCube, and IceTop    at $\sim 1$~PeV \citep{IceCube:2024pnx}. The alignment may also explain the observed deviation  from the direction of the local magnetic field above 1~PeV \citep{Ahlers:2016njd}, although anisotropy measurements at these energies remain subject to large uncertainties.

The transport of cosmic rays in the Galactic magnetic field is believed to be diffusive at least up to $10^{16-17}$~eV. In this work, we have adopted an anisotropic diffusion model with $D_\perp = 0.1 \,D_\parallel$. Variations in the diffusion coefficient, the perpendicular-to-parallel ratio, the Galactic magnetic field model, or even the transport regime itself would modify the residence time of PeV cosmic rays both near their sources and in the Milky Way more generally \citep{2017JCAP...06..046M}.  However, the correlation between the smoothness of the cosmic-ray density and the relative contribution of individual sources, and their dependence on the source birth rate, are not likely to change significantly.

We did not include heavier nuclei in our simulations because this study focuses on interpreting the proton knee. Models invoking rigidity-dependent spectral steepening, referred to as the Peters cycle \citep{Peters:1961mxb,2005JPhG...31R..95H}, can broadly account for the cosmic-ray composition measurements despite remaining challenges  \citep{Gaisser:2013bla, Prevotat:2025ktr}. In  a simplified scenario where diffusion  depends only on the rigidity, our proton-based study  would also apply to heavier nuclei. However, differences in the abundance among various source classes and the effects of the spallation of nuclei \citep{2012JCAP...01..010B} introduce additional complexity into the model. Although helium and heavier nuclei constitute a significant fraction of the  cosmic-ray spectrum, their contribution to the Galactic diffuse gamma-ray emission is considerably smaller, at the level of $\sim 30-40\%$ of the proton contribution at a  nucleon  energy of 1~PeV \citep{Fang:2023azx,Castro:2025wgf}.   

The model we presented in Section~\ref{sec:cygnus} does not include PeVatrons in the Southern sky such as the Vela supernova remnant \citep{Ahlers:2016njd}. It also neglects contributions from leptonic sources, unresolved sources, and spatially dependent diffusion, all of which may modify the longitudinal distribution of diffuse gamma-ray emission, particularly at lower energies. A more complicated model could incorporate these additional components and be constrained by full-sky and Southern-sky diffuse emission observations by IceCube and SWGO \citep{SWGO:2025taj}.

\vspace{2em}

\begin{acknowledgments}
K.F. acknowledges support from the National Science Foundation (PHY-2238916, PHY-2514194) and the Sloan Research Fellowship. This work was supported by a grant from the Simons Foundation (00001470, KF). The research of F.H was also supported in part by the U.S. National Science Foundation under grants~PHY-2209445 and OPP-2042807.
\end{acknowledgments}

\bibliography{ref}

\input{appendix}

\end{document}

%% file: appendix.tex

\appendix

\section{Numerical Setup for Cosmic-ray Propagation}\label{appendix:setup}
We compute the cosmic ray propagation in the Galactic magnetic field using the CRPropa~3.2 package \citep{AlvesBatista:2022vem}. In this Monte Carlo approach, cosmic-ray transport with respect to the structure of the Galactic magnetic field is accounted for  through an anisotropic diffusion tensor. We adopt a parallel diffusion coefficient \citep{Strong:1998pw}, $D_\parallel = 6.1\times 10^{28} \left( {R}/{4\,\rm GV}\right)^{1/3}\,\rm cm^2\,s^{-1}$,
where $R = E/Z$ is the rigidity of a cosmic-ray particle with energy $E$ and charge $Z$. We set the perpendicular diffusion  $D_\perp = 0.1 D_\parallel$. 

We calculate the trajectories of 50 million pseudo particles from $N_s$ sources, propagating them forward in time till the maximum trajectory reaches the source age ($c \,t_{\rm age}$) or a particle escapes from a cylindrical simulation boundary at $r_{\rm max} = 20$~kpc and $|z|_{\rm max} = 10$~kpc. We have verified that the evolution of cosmic-ray density near the solar neighborhood barely changes when $|z|_{\rm max}$ is increased beyond 10~kpc. We use adaptive steps with minimal and maximal step size of $1$~pc$/c$ and  $1$~kpc$/c$, respectively, and a precision of $\xi = 10^{-4}$. Particle positions are recorded at 10  points in linear  time bins in the study of Section~\ref{sec:birthrates} or at 20-100 points in logarithmic time bins in the analysis of Section~\ref{sec:testSource} and Section~\ref{sec:cygnus}. The simulated particle energy distribution follows $dN/dE\propto E^{-1}$ from 3~TeV to 30~PeV, and is latter scaled to the desired injection spectrum.

For the Galactic magnetic field, we adopt the JF12 field model \citep{2012ApJ...757...14J, 2012ApJ...761L..11J}, which consists  of a large-scale regular field, a  striated random component, and a small-scale, random turbulent field.

Proton-proton interactions are neglected in the calculations presented in Section~\ref{sec:birthrates}. This approximation is justified because the proton interaction timescale in the interstellar medium is of order $1 / (n_{\rm ISM} \sigma_{\rm pp}c) \approx 15\,(n_{\rm ISM} / 1\,{\rm cm^{-3}})^{-1}\,\rm Myr$, adopting an  inelastic cross section $\sigma_{\rm pp}\approx 70$~mb at PeV energies. Because most PeV protons escape  the Galaxy within tens of Myrs after injection, such interactions has minimal impact on their propagation.

The simulations are performed on a computing cluster and takes approximately 6,000 to 30,000 CPU hours per simulation set.

\section{Effects of source duration}\label{appendix:sourceLifetime}

\begin{figure*}
    \centering
   \includegraphics[width=1.06\textwidth]{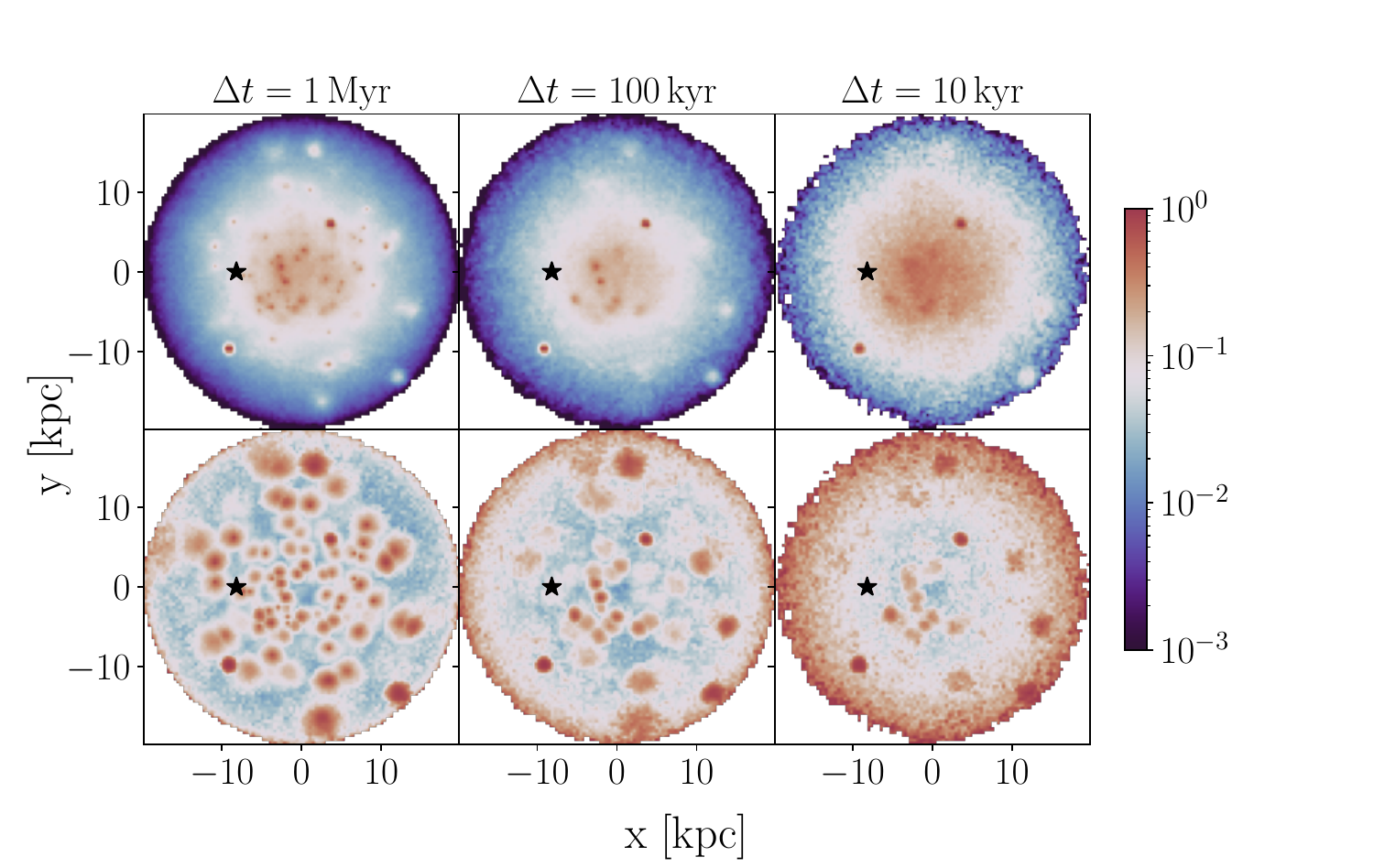}
   \caption{
    \label{fig:1PeV_1e4} Same as Figure~\ref{fig:1PeV}, but with the source birth rate fixed at $f_{\rm SN}=1\%$ and  the source duration varied as $\Delta t =$~1~Myr, 100~kyr, and 10~kyr.  } 
\end{figure*} 

Figure~\ref{fig:1PeV_1e4} illustrates how the source duration affects the spatial distribution of PeV cosmic rays and the relative contribution from individual sources. We fix the source birth rate at $f_{\rm SN} = 1\%$ and vary the source duration $\Delta t$. Compared to the average leading-source contribution of 11.3\%  in the case of $\Delta t  = 100$~kyr (see Section~\ref{sec:birthrates}), this fraction changes to 9.8\%  and  18.1\% when $\Delta t$ is increased or decreased by an order of magnitude, respectively.

A longer $\Delta t$ also allows more sources to dominate the PeV cosmic-ray flux of their local vicinities. The spatial extent of  regions dominated by a  single source, on the other hand, appears to be primarily determined by the source birth rate. This extent is similar across all cases with $f_{\rm SN}=1\%$, but is larger and smaller  than those in the $f_{\rm SN}=10\%$ and 0.1\% cases in Figure~\ref{fig:1PeV}, respectively.

\begin{figure}
    \centering
   \includegraphics[width=0.49\textwidth]{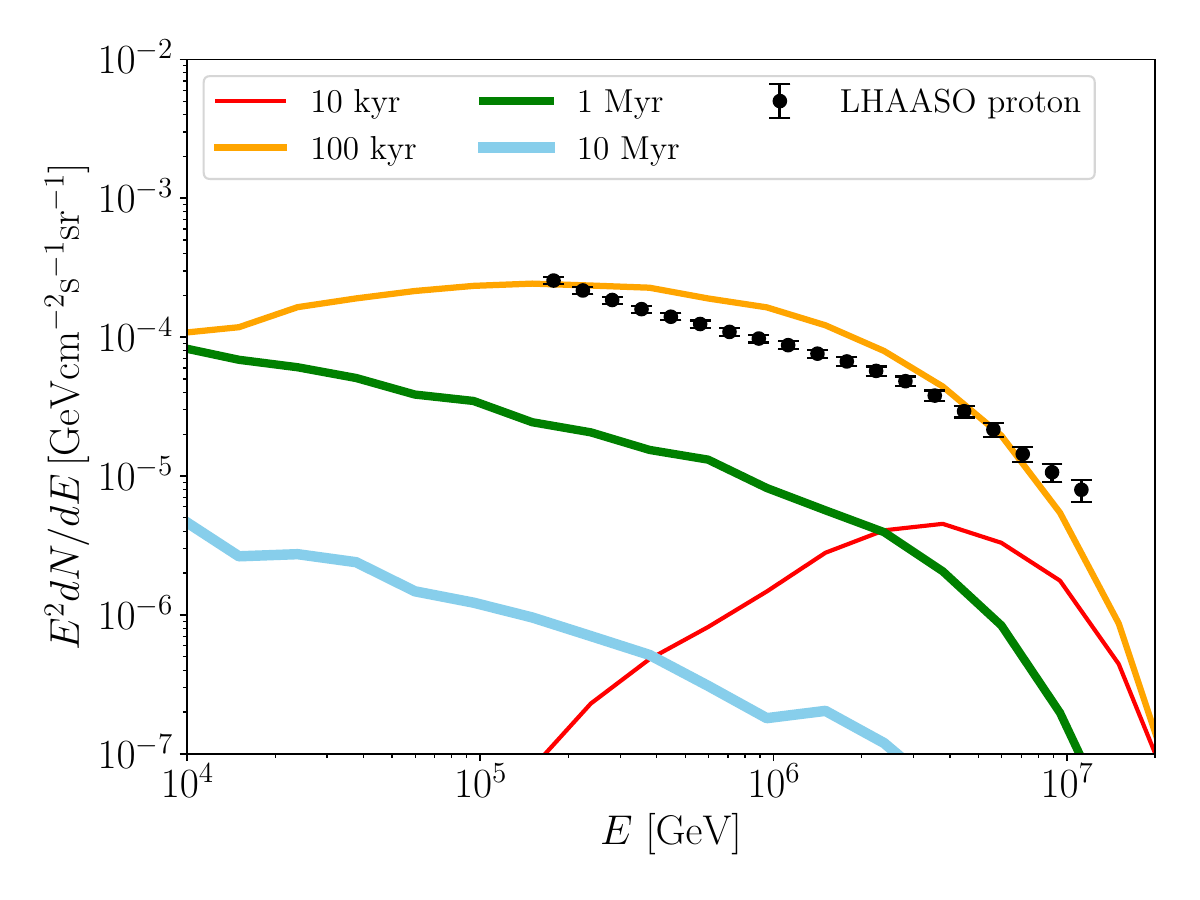}
   \caption{
    \label{fig:spectrum_source_dt} Same as Figure~\ref{fig:spectrum_source}, but with source duration fixed at $\Delta t=10$~kyr for fiducial sources with ages of $t_{\rm age} =$ 10~kyr, 100~kyr, 1~Myr, and 10~Myr.  } 
\end{figure}

Figure~\ref{fig:spectrum_source_dt} presents the local cosmic-ray spectrum from a fiducial source located $d=1$~kpc away as in Section~\ref{sec:testSource}, except that the source is assumed to emit  cosmic rays only at the beginning of its lifetime, over a duration of $\Delta t = 10$~kyr. For a recent source with $t_{\rm age}=100$~kyr, the resulting local cosmic-ray flux is similar to that in Figure~\ref{fig:spectrum_source}. When $\Delta t = t_{\rm age}$, early-injected particles have already passed the observer, while late-injected ones have not yet arrived; when $\Delta t = 10$~kyr, a substantial fraction of particles has already passed. As a result, the cosmic-ray density near the observer is comparable in both cases.

The effect of a short source duration is more prominent for source ages of 1 and 10~Myr. In these cases, since  $\Delta t \ll t_{\rm age}$, particle injection is effectively instantaneous, and the cosmic-ray density in the disk  can be approximated by  the leading-order term in the solution to the diffusion equation, 
\begin{equation}
    n(\vec{x}, E, t) \propto \frac{1}{4\pi D(E) H t} e^{-r^2 / 4D(E) t} e^{-t / \tau_{\rm esc}}
\end{equation}
where $r = |\vec{x} - \vec{x}_{\rm source}|$ is the distance between the observer and the source, and $\tau_{\rm esc}= 4 H^2 / \pi^2 D \sim H^2/2D$ is the characteristic escape time from a Galactic halo extending to a height $H$ above the plane \citep{Cowsik:2025xeg}. In both cases, cosmic-ray leakage reduces the remaining flux to a level too low to account for the knee. Gamma-ray emission by cosmic rays from these relic sources is expected to be diffuse-like and not to reveal the central source. 

In the case of an instantaneous injection,  the cosmic-ray density at a fixed time, $n(r)\propto e^{-r^2 / 4Dt}$, leads to a dipole amplitude of $\delta\approx  3D /c \, (\nabla n/   n) \sim   d / c\,t = 3\times 10^{-3}\left(d/1\,{\rm kpc}\right)\left(t/1\,{\rm Myr}\right)^{-1}$. The dipole amplitude increases for sources that are younger or more distant.

\section{Spatial variation of the cosmic-ray spectral index}\label{appendix:index}

\begin{figure*}
    \centering
   \includegraphics[width=0.89 \textwidth]{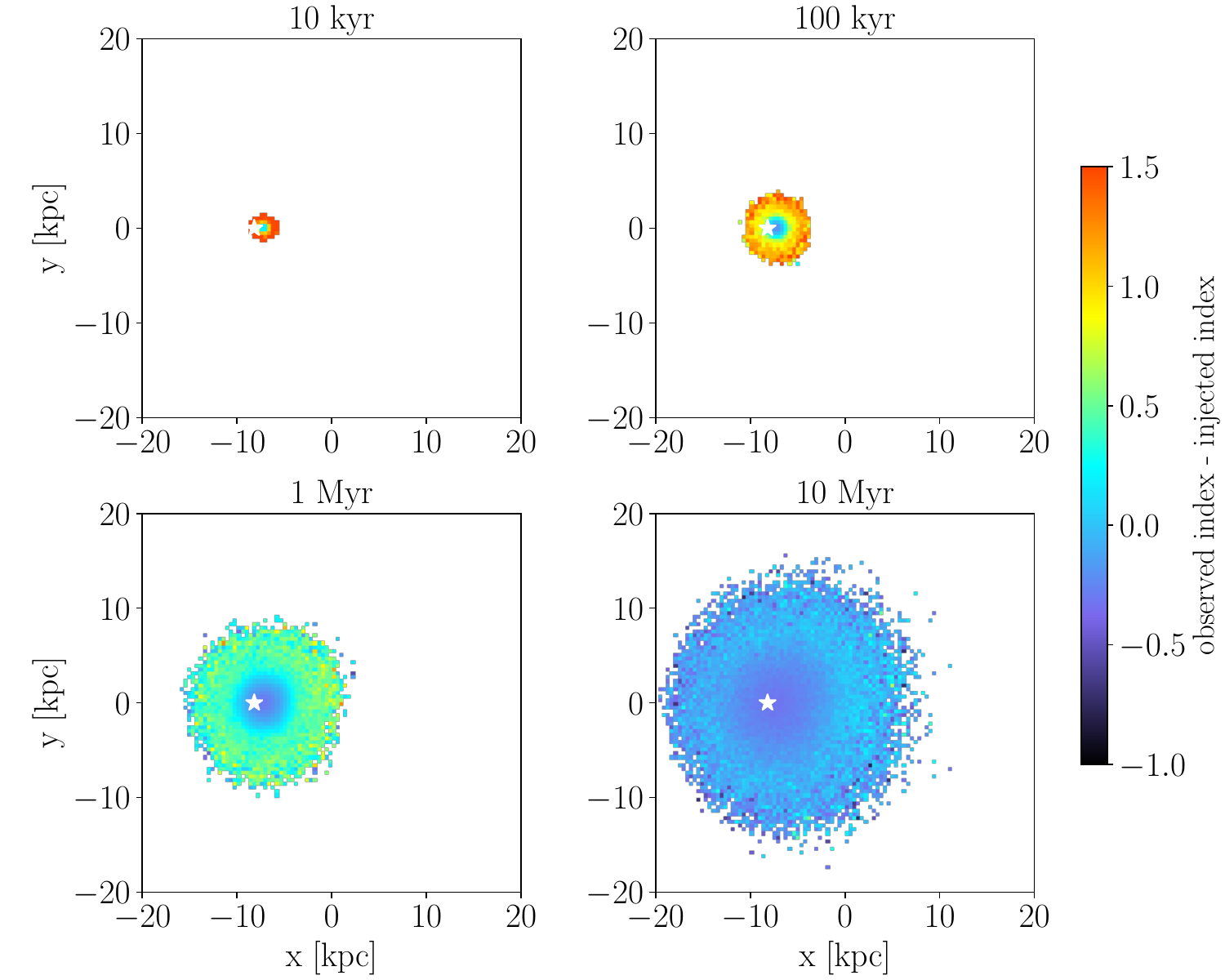}

    \caption{
    \label{fig:index_source} Difference in the injected and observed spectral index, indicated by the color, at various positions in the disk, of cosmic rays from a source located at 1~kpc from the sun in the direction of the Galactic center, emitting cosmic rays at a constant rate over  durations of 10~kyr (top left), 100~kyr (top right), 1~Myr (bottom left), and 10~Myr (bottom right). All plots are face-on views for a slice at $z=0$. Harder (softer) spectra relative to the injection appear toward the red (black) end of the color scale.   } 
\end{figure*} 

To evaluate the impact of an individual source on the cosmic-ray spectrum in its vicinity,  we  propagate cosmic rays from four fiducial sources with ages ranging from 10~kyr to 10~Myr, as in Section~\ref{sec:testSource}. The injected cosmic rays follow a single power law with index $-2$,  $dN/dE\propto E^{-2}$. We then evaluate the cosmic-ray spectrum in the Galactic disk by fitting a power-law, $dN/dE\propto E^{\alpha}$, to the propagated cosmic rays. Figure~\ref{fig:index_source} shows the deviation of the propagated index from the injection value, $\alpha+2$, as a function of positions in the disk.

Figure~\ref{fig:index_source} illustrates how  the cosmic-ray spectrum  depends on both the distance from the source and the time elapsed since injection. 
It also  explains the spectra shown in Figure~\ref{fig:spectrum_source}, which   apply only to the solar position.

\section{Secondary Production}\label{appendix:secondary}

We compute the intensity of gamma-ray emission from diffuse cosmic rays  using the \texttt{HERMES} code \citep{HERMES}. The calculation accounts for the cosmic-ray interaction with neutral (HI) and molecular hydrogen (H$_2$) using the proton-proton interaction cross section from \citet{2006PhRvD..74c4018K}. Following \citet{HERMES}, we assume  that the ISM gas is composed of hydrogen and helium nuclei with a uniform abundance ratio $n_{\rm He} / n_{\rm H} = 0.1$. The resulting gamma-ray emission is computed as the sum of the products of proton cosmic rays interacting with both hydrogen and helium gas. The H$_2$ gas density is  traced by the CO emission using an CO-H$_2$ conversion factor, $X_{\rm CO} = 1\times 10^{20}\,\rm cm^{-2}\,K^{-1}/(km\,s^{-1})$. Our gamma-ray calculations used a linear 3D grid with a spacing of $0.2$~kpc in x and y directions, spanning -35~kpc to 35~kpc, and a spacing of $0.17$~kpc in z direction, spanning from -5~kpc to 5~kpc.